# Design and Analysis of a Sample-and-Hold CMOS Electrochemical Sensor for Aptamer-based Therapeutic Drug Monitoring

Jun-Chau Chien, *Member, IEEE,* Sam W. Baker, H. Tom Soh, Amin Arbabian, *Senior Member, IEEE*

*Abstract*—In this paper, we present the design and the analysis of an electrochemical circuit for measuring the concentrations of therapeutic drugs using structure-switching aptamers. Aptamers are single-stranded nucleic acids, whose sequence is selected to exhibit high affinity and specificity toward a molecular target, and change its conformation upon binding. This property, when coupled with a redox reporter and electrochemical detection, enables reagent-free biosensing with a sub-minute temporal resolution for *in vivo* therapeutic drug monitoring. Specifically, we design a chronoamperometry-based electrochemical circuit that measures the direct changes in the electron transfer (ET) kinetics of a methylene blue reporter conjugated at the distal-end of the aptamer. To overcome the high-frequency noise amplification issue when interfacing with a large-size (> 0.25 mm²) implantable electrode, we present a sample-and-hold (S/H) circuit technique in which the desired electrode potentials are held onto noiseless capacitors during the recording of the redox currents. This allows disconnecting the feedback amplifiers to avoid its noise injection while reducing the total power consumption. A prototype circuit implemented in 65-nm CMOS demonstrates a cell-capacitance-insensitive input-referred noise (IRN) current of 15.2 pA$_{rms}$ at a 2.5-kHz filtering bandwidth. Tested in human whole blood samples, changes in the ET kinetics from the redox-labeled aminoglycoside aptamers at different kanamycin concentrations are measured from the recorded current waveforms. By employing principal component analysis (PCA) to compensate for the sampling errors, a detection limit (SNR = 1) of 3.1 µM under 1-sec acquisition is achieved at 0.22-mW power consumption.

This work was supported by Stanford Precision Health and Integrated Diagnostic Center (PHIND), National Institute of Health (NIH) SPARC Program (OT2OD025342), NIH NIBIB (R01EB025867), National Science Foundation (NSF) CAREER Award (ECCS-1454107), and the Chan-Zuckerberg Biohub.

J.-C. Chien and A. Arbabian are with the Department of Electrical Engineering, Stanford University, Stanford, CA 94305 USA (e-mail: jcchien@stanford.edu, arbabian@stanford.edu).

S. Baker is with the Department of Comparative Medicine, Stanford University, Stanford, CA 94305 USA (e-mail: sambaker@stanford.edu).

H. T. Soh is with the Department of Radiology and the Department of Electrical Engineering, Stanford University, Stanford, CA 94305 USA (e-mail: tsoh@stanford.edu).

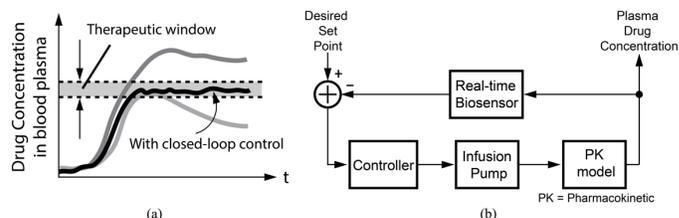

Fig. 1. (a) Real-time therapeutic drug monitoring enables precision drug dosing for optimal treatment outcome. (b) A generic closed-loop drug control system.

## I. INTRODUCTION

Current clinical standard for drug dosing relies on physical parameters, such as age, gender, body weight, and body surface area. Unfortunately, these approaches do not account for individual differences in pharmacokinetics (PK), which describes the absorption, distribution, metabolic, and excretion rate of drugs in the body [1–3]. As PK between individuals can exhibits difference as large as tenfold [4], there is great interest in developing the technology for monitoring and controlling drug concentrations *in vivo* for optimal therapeutic outcomes at minimal toxicity (Fig. 1(a)) [5–7]. Such control is particularly important for drugs with narrow "therapeutic windows" wherein underdosing results in low efficacy and over-dosing can cause acute injury to organs [8–10]. Today, therapeutic drug monitoring (TDM) requires multiple venous blood draws and samples are analyzed using immunoassays [11–12] or high-performance liquid chroma-tography/mass-spectroscopy (HPLC/MS) [13–14]. These assays require *hours* of processing which is insufficient for optimal drug dosing. Thus, there is a pressing need for continuous biosensor that can measure drug levels in real-time. Importantly, such a sensor would enable the development of closed-loop systems for automatically delivering optimal doses of drugs to individual patients regardless of their PK profile. (Fig. 1(b)).

To this end, our group and others have previously utilized *aptamers* to achieve continuous detection of drugs *in vivo* [15–21]. Aptamers are "synthetic antibodies" composed of nucleic acids that can specifically bind to the target analytes in complex samples such as the whole blood [22–24]. Importantly, they can be engineered into "aptamer-switches" that undergo *structure-switching* upon target binding in a reversible manner. By conjugating electroactive reporters to the aptamers, the changes in their structure (thus the analyte concentration) can be detected electrochemically. As sample preparation is not needed, aptamer switches are capable of continuous monitoring of biomolecules *in vivo*.

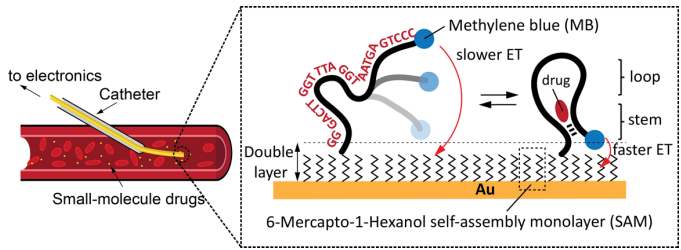

Fig. 2. The operation of structure-switching aptamer with and without the presence of the drug and its use in in vivo monitoring.

Nevertheless, the current generation of aptamer-based real-time biosensors requires either a continuous drawing of the subject's blood [15] or wired connection to an implanted device [16], and both systems are only suitable for non-ambulatory patients. To overcome these limitations, we have previously reported a miniaturized CMOS system for implantation [25]. However, this system consumed milliwatts of power, dominated by the signal-to-noise ratio (SNR) requirement in the sensing circuits when interfacing with large-area-electrodes for *in vivo* monitoring. Moreover, the readout utilizes square-wave voltammetry (SWV) with a long acquisition time (~8 sec), leading to an excessively high energy consumption (> 50 mJ/ sample). This motivates us to develop circuit techniques that can overcome such a noise-power trade-off

Several electrochemical-sensing current-readout circuits have been developed for *in vitro* DNA and biomolecule analysis [26–31], *in vivo* glucose and lactate monitoring [32–34], and sub-second neurotransmitter detection [35–38]. In these implementation, transimpedance amplifier (TIA) and current conveyor (CC) are two popular circuits topologies for reading small currents. As both employ transconductance amplifiers (OTAs) in feedback to establish proper electrode potentials, it is inevitable to couple the OTA noise to the impedance of the electrochemical cell, resulting in noise amplification when measured at high bandwidth (> 1 kHz). This is especially problematic in our specific applications.

In this work, we present a sample-and-hold (S/H) technique in chronoamperometry (CA) to overcome the noise/power trade-off when detecting aptamer conformation switching [39]. By holding the desired potentials onto the electrodes using capacitors, we successfully record, for the *first* time, the changes in the *electron-transfer (ET) kinetics* of the redox reporters without OTAs. Such an open-loop scheme enables simultaneous reduction in both the circuit noise and the power consumption. As a proof-of-concept, we used our circuit with an aminoglycoside aptamer [15] to measure kanamycin concentrations and achieved limit-of-detection (LoD) that is on-par with benchtop laboratory instruments. Built upon [39], this paper provides more details regarding the signal transduction mechanism and the analysis that assists the mapping between the electronics noise and the measurement uncertainty in molecular concentration. The issue of sampling error will be discussed, and a compensation technique is presented.

This paper is organized as follows. Section II describes the sensing mechanism and the detection schemes for the aptamer switches. The proposed circuit architecture, signal and noise analysis, as well as its implementation are introduced in Section III. Section IV presents the experimental results, covering both the electronics characterization and assay experiemtns. Conclusions are presented in Section V.

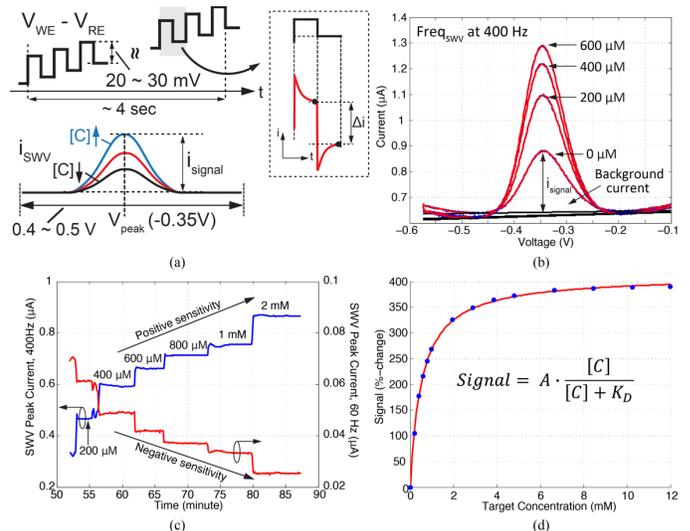

Fig. 3. (a) SWV operation. (b) SWV voltammograms at 400 Hz at different kanamycin concentrations. (c) Measured time-series at 400 and 60 Hz SWV frequencies. (d) Fitting of the binding curve with Langmuir isotherm model.

## II. APTAMER SWITCHES

### A. Sensing Mechanism

As described above, our group has utilized structure-switching aptamers to achieve real-time detection of small-molecule drugs *in vivo* [15][17]. Using the aminoglycoside aptamer as an example (Fig. 2), the binding to the target molecules promotes a more thermodynamically stable *stem-loop* structure (folded) instead of a linear and flexible structure (unfolded). By conjugating the aptamers with a methylene blue (MB, $C_{16}H_{18}ClN_3S$) reporter at the distal end of the DNA, such a conformation change can be detected by measuring the differences in the electron-transfer kinetics of MB through its sensitivity to the diffusion distance versus the underlying electrode [40–41]. The highest ET kinetics occurs when the MB is closer to the electrode, resulting in a larger electrical current when measured in voltammetry. In this way, higher drug concentration will cause more aptamers being switched into the stem-loop configuration, and thus a larger accumulated signal. In our system, these aptamers are immobilized at the tip of a sensing probe, which can be implanted directly into a vein for measuring drug concentration in blood plasma.

### B. Square-wave Voltammetry

The states of the aptamers are generally measured using square-wave voltammetry (SWV; Fig. 3(a)) [15–16, 40–41]. Unlike cyclic voltammetry (CV), SWV measures the ET kinetics through continuous modulation of the MB states using a small-amplitude square-wave superimposed on a stepping voltage established between a reference (RE) and a working electrodes (WE). This effectively separates the non-Faradaic currents (charging and discharging currents of the interfacial double-layer capacitance) from the desirable Faradaic components (from redox reactions), relaxing the dynamic range requirement and enhancing the sensitivity [40]. Fig. 3(b) depicts example measurements at different kanamycin concentrations (in buffer) using a commercial potentiostat (Palmsens EmStat3 Blue). To mitigate sensor drifts, we employ kinetic differential measurement (KDM) technique [15], which takes advantage of the sensitivity difference measured at two SWV frequencies (e.g. 400 and 60 Hz) to reject common-mode drift while preserving the desired signal (Fig. 3(c)). The molecular limit-of-detection (LoD), defined at SNR = 1, is 1.5 µM for amino-glycoside aptamer with a dissociation constant ($K_D$) of 0.5 mM for kanamycin (extracted from the binding curve in Fig. 3(d) with a Langmuir isotherm model). These numbers will be used later for benchmarking electronics performance. One critical drawback of SWV is the long scanning time (~8 sec per acquisition), which translates to significantly higher energy consumption per readout sample.

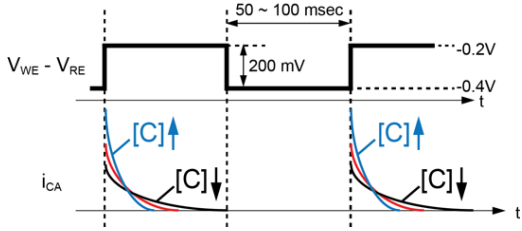

Fig. 4. Chronoamperometry operation.

Table I. A comparison between SWV and CA.

| | Square-wave Voltammetry (SWV) | Chronoamperometry (CA) |
|---|---|---|
| Kinetic measurements | Indirect | Direct |
| Temporal resolution | 2 sec per scan (< 0.5 Hz) | < 0.2 sec (> 5 Hz) |
| Current drift compensation | Requires 2 scans at different SWV frequency | Direct time-constant extraction |
| Curve-fitting error | Sensitive to background level | Less sensitive |
| $V_{peak}$ tracking | Yes | No |

### C. Chronoamperometry

Though SWV has been the mainstream for electrochemical-based aptamer sensing, chronoamperometry (CA) has recently been re-investigated for redox-coupled DNA detection [42–43]. As shown in Fig. 4, chronoamperometry operates by pulsing the electrode potential ($V_{WE} - V_{RE}$) between two largely-separated voltage levels ($V_1 - V_2 > 200$ mV) to modulate the energy states of the redox molecules from the fully-reduced toward oxidation (and vice versa). The kinetics of the electron transfer is then probed *directly* through measurements of the decaying current transients. Chronoamperometry offers three main advantages over SWV: (1) higher temporal resolution due to the absence of potential scanning; (2) reduced sensitivity to the drifts in the absolute current (rate measurements); and (3) minimal influence from background interference due to a limited voltage-scanning range (Table I). More importantly, its operational simplicity offers an opportunity to incorporate circuit technique for simultaneous power and noise reduction, as will be discussed in Section III-C. Nevertheless, CA is more sensitive to the instability of the RE potential because any drift in the redox potential will remain undetectable and leads to signal variation. Such an issue can be solved with a hybrid scheme using SWV (or cyclic voltammetry) to frequently calibrate and re-adjust $V_1$ and $V_2$ [37]. In our device, the oxidation (reduction) potential of the MB measured from our device is located at -0.3V (-0.4V), and hence $V_1$ and $V_2$ are selected as -0.2V (-0.3V) and -0.4V (-0.5V).

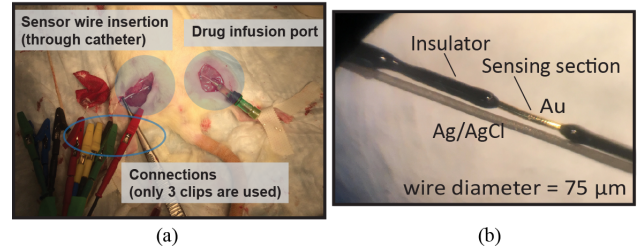

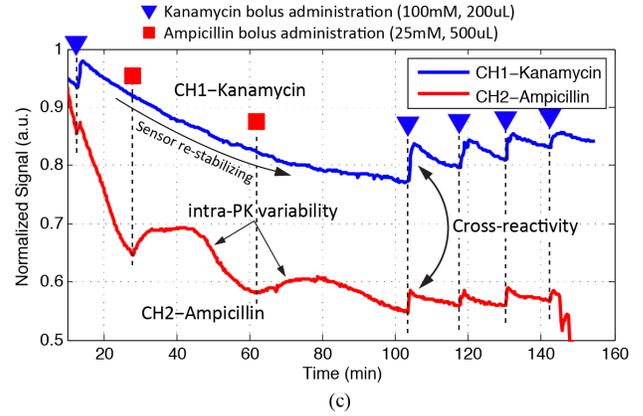

Fig. 5. (a) in vivo demonstration with a rodent model. (b) The photo of the implanted probe. (c) Measured responses of two aptamers.

### D. Measurement Uncertainty

It is crucial to study factors that impact the performance of the biosensor performance *in vivo*. Generally, the uncertainty in the assay ($e^2_{assay}$) is summarized as [44]:

$$e^2_{assay} = e^2_{n,electronics} + e^2_{n,MSN} + e^2_{background}, \quad (1)$$

where $e^2_{n,electronics}$ refers to the noise from the measurement electronics, $e^2_{n,MSN}$ represents the shot noise induced by both the molecular binding and the electron transfer in the redox reaction, and $e^2_{background}$ includes all other uncertainty such as drifts, sensitivity degradation, non-specific binding, and curve-fitting errors. We carried out both *in vitro* (using flowing human whole blood) and *in vivo* (using an anesthetized rodent) studies to monitor aptamer responses under changing drug concentration. Fig. 5 demonstrates an example of *in vivo* setup, our implantable device (which will be used later when testing the proposed sensor circuits), and the measurement results (in SWV). We found that $e^2_{background}$ occurs at a longer time scale (~ hours) and can be corrected with *in vitro* calibration, similar to the approach used in continuous glucose monitoring (CGM) system [45], and $e^2_{n,MSN}$ is negligible due to relatively high target concentration in our drug-dosing application. On the other hand, $e^2_{n,electronics}$ plays the dominant role in determining the sample-to-sample variance. This is indeed the case in our earlier CMOS implementation [25] where we measured an LoD of only 18 µM ($\approx 0.036 K_D$).

Note that $V_{peak}$ in the acquired SWV curves is shifted by ~15 mV in the 3hr *in vivo* experiments due to the drift in the pseudo-RE. We adjusted the upper and lower bounds of the scanned voltages to re-center $V_{peak}$ for measurement consistency. As mentioned earlier, such an adjustment is particularly important in CA measurements.

### E. Implantable Probe

To fit into the vein of a rodent, we prepare an implantable probe (Fig. 5(b)) consisting of a bundle of a gold (Au, the WE) and a silver chloride (Ag/AgCl, the RE) wires, all at 75-μm diameter. The sensing area on the gold electrode is defined using heat-shrinkable polymer tubing. Here we employed a two-electrode electrochemical cell instead of a standard three-electrode setup. This is possible because the electrolyte resistance between the RE and WE is sufficiently small such that the IR drop between them is negligible. The aptamers are immobilized onto the oxidized gold wire through thiol bonds (S-H). The device is implanted through the femoral vein using a 20-gauge catheter. The system can be scaled up by bundling extra gold wires functionalized with different aptamers to monitor more than one type of drugs. Fig. 5(c) demonstrates the results of a multiplexed drug detection with bolus administration of two drugs (kanamycin and ampicillin) intravenously at different time instants. The sequence of aptamers for detection ampicillin is adapted from [46].

It is critical to point out the importance of aptamer packing density on the electrode. On one hand, excessively high packing density induces undesired inter-molecular interactions among the aptamers and cause sensitivity degradation. On the other hand, excessively low packing density produces small signals. In general, we prepare the aptamers at a surface density of 1% which produces an optimal signal [47]. This translates to a 10-nm average spacing among each aptamer assuming each one occupies a real estate of 1 nm$^2$, or $10^{12}$/cm$^2$. To cope with such a low surface density, we have chosen a relatively large area (> 0.25 mm$^2$) in our implantable sensing probe to achieve a sufficiently large signal without the need of surface roughening [48] and to avoid the issue of irreversible oxygen reduction encountered when using a small-area electrode [49]. Inevitably, such a large cell capacitance will have system design implications and causes significant noise amplification, which will be discussed in Section III-A.

## III. APTAMER-SENSING CIRCUITS

### A. Signal Analysis and Design Specification

In this section, we focus on understanding the impact of electronics noise to the LoD of the measured drug concentration in chronoamperometry. Before starting, it is important to mention that the extraction of the concentration information from the decaying current waveforms involves curve-fitting technique. Such an "averaging" effect on the electronics noise must not be neglected otherwise the design specification will be overly conservative. Our approach is detailed in the next paragraph.

To simplify the noise analysis, we model the ET kinetics using an exponential function with a time constant $\tau_0$[1] (Fig. 6):

$$i_{redox}(t) = \frac{\alpha Q_T}{\tau_0} e^{-\frac{t}{\tau_0}}. \qquad (2)$$

[1] Deriving the actual current response under overpotential activation requires solving the mass transport kinetics using Fick's law of diffusion and Butler-Volmer equation [50–51], and this often requires numerical simulation based on the electrode geometry [41].

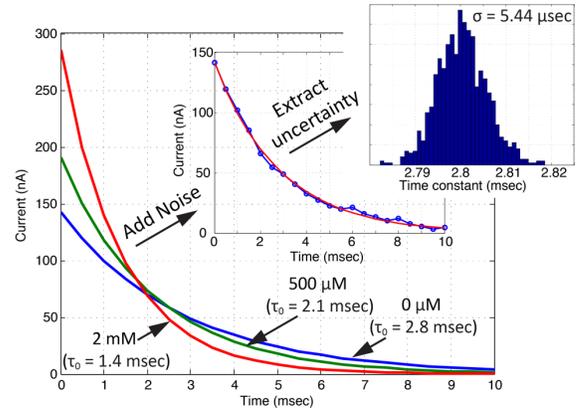

Fig. 6. Monte Carlo simulation for studying impact of electronics noise.

Table II. Design Specification

| Biosensor | |
|---|---|
| Electrode area | 0.25 mm$^2$ (Au) |
| Probe density | ~1% ($10^{12}$/cm$^2$) |
| Self-assembly monolayer | 6-mercapto-1-hexanol HS(CH$_2$)$_6$OH |
| Electrode capacitance $C_{DL}$ | w/o passivation: ~60 nF w/ passivation: ~10 nF |
| Drug Target | Kanamycin monosulfate C$_{18}$H$_{36}$N$_4$O$_{11}$ · H$_2$SO$_4$ (Aminoglycoside) |
| Aptamer sequence | 5'-GGGACTTGGTTTA GGTAATGAGTCCC-3' |
| Redox label | Methylene blue |
| Redox kinetics ($\tau_0$) | 1.2 ~ 3 msec (1:2.5) |
| Molecular resolution | < 10 μM ($K_D$ = 0.5 mM) |
| Chronoamperometric Sensor | |
| Dynamic range | 60 dB |
| Bandwidth | 2.5 kHz |
| Electronics noise | < 250 pA$_{rms}$ |
| Potential ranges | -0.4 ~ -0.2V (adjustable) |
| Acquisition rate | 5 ~ 10 Hz |

In (2), $Q_T$ represents the total charge when all the MB electrons (two electrons per molecule) are fully transferred during the redox reaction, and $\alpha$ accounts for the percentage of the MB that participates in the electron transferring. The weighting term ensures that the total amount of transferred charges under varying $\tau_0$ remains constant. Next, we perform Monte Carlo simulation with Gaussian white noise added to the time series and the curve is fitted (*fit* function in Matlab) to extract the kinetical time constant. The procedure is repeated multiple times at different noise magnitude, and the uncertainty (one standard deviation) in the time constant is calculated. Finally, we map the electronics noise to the molecular concentration assuming a known sensitivity ($S = \Delta\tau_0 / \Delta[C]$).

Fig. 6 shows the calculated current for a 0.25-mm$^2$ (= 500 μm × 500 μm) electrode at 1% packing density. Here we assume a time-constant ratio of 2:1 at high and low target concentration based on our earlier experiments. According to our simulation, an additive noise magnitude of < 0.25nA$_{rms}$ (sampled at 2.5 kHz over 10 msec) is needed to ensure that the concentration uncertainty is kept below 0.01$K_D$. The readout IC requires approximately 10 bits of dynamic range with the maximum signal determined by the fastest electron transfer kinetics (k ~ 440 1/sec) [40]. We summarize the design specification for both the biosensor and the electrochemical-sensing interface circuits in Table II.

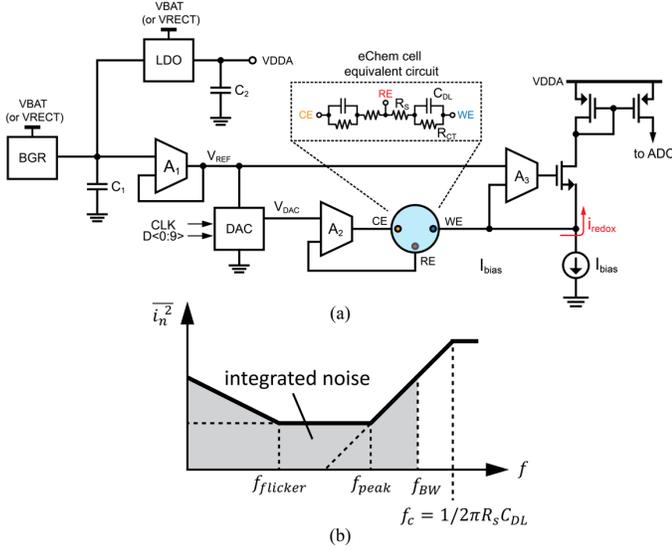

Fig. 7. (a) The system block diagram of a conventional electrochemical-sensing circuit. (b) PSD illustration of noise peaking.

The electron transfer kinetics is also impacted by the use of the self-assembly monolayer (SAM) for passivating the sensor surface. Such passivation is necessary to block the undesirable redox reaction from the electroactive biomolecules presented in the biofluids, including oxygen, uric acids, ascorbic acids, and etc. In our *in vivo* assay, we employ 6-mercapto-1-hexanol (MCH; $C_6H_{14}OS$, a hydroxyl-terminated six-carbon chain) to serve this purpose and to balance between electron transfer kinetics and the stability of the SAM [52]. The presence of the passivation also affects the double-layer capacitance at the electrode-electrolyte interface. The measured double layer capacitance ($C_{DL}$) exhibits a unit capacitance of 36 and 215 $nF/mm^2$ with and without the passivation, respectively. The estimated $C_{DL}$ for our implanted electrode is also included in Table II. In our circuit design, we have started with $C_{DL}$ = 100 nF as the worst-case scenario (assuming no passivation and taking into account variation in the sensor surface area).

### B. Electrochemical-sensing Interface Circuits

Fig. 7 shows the conventional electrochemical readout circuit [25]. It consists of a bandgap reference (BGR) voltage generator, a reference buffer ($A_1$), an R-2R digital-to-analog converter (DAC), a control amplifier ($A_2$), a current conveyor (CC) with feedback amplifier ($A_3$), and an analog-to-digital converter (ADC). The total noise current, when referred to the input of the current conveyor (or WE), is the power sum of the electrochemical-cell-dependent noise and the additive noise from the current conveyor:

$$\overline{i_{n,WE}^2} = \overline{i_{n,1}^2} + \overline{i_{n,2}^2}, \quad (3)$$

$$\overline{i_{n,1}^2} = \frac{\overline{v_{n,BGR}^2} + \overline{v_{n,DAC}^2} + \overline{v_{n,A1}^2} + \overline{v_{n,A2}^2} + \overline{v_{n,A3}^2}}{|Z_{cell}|^2}, \quad (4)$$

$$\overline{i_{n,2}^2} = \overline{i_{n,Ibias}^2} + \overline{i_{n,M1}^2} + \overline{i_{n,M2}^2} + \frac{1}{N_{mirror}^2}\overline{i_{n,ADC}^2}. \quad (5)$$

In (4), $Z_{cell}$ (= $R_s + 1/j\omega C_{DL}$) models the electrical impedance of the electrochemical cell, and $\overline{v_{n,BGR}^2}$, $\overline{v_{n,DAC}^2}$, $\overline{v_{n,A1}^2}$, $\overline{v_{n,A2}^2}$, $\overline{v_{n,A3}^2}$ are the output noise of the bandgap reference, the DAC, and the input-referred voltage noise of the amplifiers $A_{1-3}$, respectively; in (5), $N_{mirror}$ is the gain of current mirror, $\overline{i_{n,M1}^2}$, $\overline{i_{n,M2}^2}$, $\overline{i_{n,Ibias}^2}$, and $\overline{i_{n,ADC}^2}$ represent the noise current from $M_1$, $M_2$, $I_{bias}$, and the ADC, respectively. According to (4), a simultaneous increase in the power consumption from each noise-contributing block is necessary to effectively reduce the overall noise. In addition, the reduced impedance from the large cell capacitance ($C_{DL}$) at frequencies above 1 kHz causes significant noise peaking and dominates the overall noise performance. In other words, there exists a noise-power trade-off when interfacing with our implantable electrodes. Fig. 7(b) shows an illustration of the power spectral density (PSD) where $\omega_{peak}$ is the peaking frequency and occurs when $\overline{i_{n,1}^2} = \overline{i_{n,2}^2}$. Assuming that the BGR noise is effectively filtered, $R_s \ll 1/\omega_{peak}C_{DL}$, and each component contributes equally, (4) and (5) are lumped to:

$$\frac{\overline{i_{n,1}^2}}{\Delta f} = N_1(\omega_{peak}C_{DL})^2 \frac{8kT\gamma}{gm_1}, \quad (6)$$

$$\frac{\overline{i_{n,1}^2}}{\Delta f} = N_2(4kT\gamma gm_2). \quad (7)$$

In (6) and (7), $gm_1$ is the transconductance of the input transistors in the amplifiers $A_{1-3}$, $gm_2$ is the transconductance of the transistors $M_1$, $M_2$, and $I_{bias}$ in the current conveyor, $k$ is the Boltzmann constant, $T$ is the absolute temperature, $\gamma$ is the technology-dependent noise factor, and $N_1$ and $N_2$ are the numbers of noise components. In (6), the output noise voltage of the R-2R DAC (with a unit resistor of $R_0$) is converted to the input-referred noise of a transistor having an equivalent gm of $\gamma/R_0$. Assuming $N_1 = N_2$, we have:

$$gm_1 \approx \frac{2}{gm_2}(\omega_{peak}C_{DL})^2. \quad (8)$$

With a $C_{DL}$ and $gm_2$ of 100 nF and 30 μS (at $I_{bias}$ = 2μA), a $gm_1$ of 26.4 mS is needed for $\omega_{peak}$ at 1 kHz. Such a gm requirement leads to excessive-high current (~2 mA per input transistor) for an optimistic 150 pA$_{rms}$ when integrated over 2.5 kHz of bandwidth (not considering the flicker noise and assume brick-wall filtering). Note that the need for high bandwidth is the main difference in power consumption requirement when compared to glucose sensors employing fixed-potential amperometry with excessive filtering (below few Hz) [32–33].

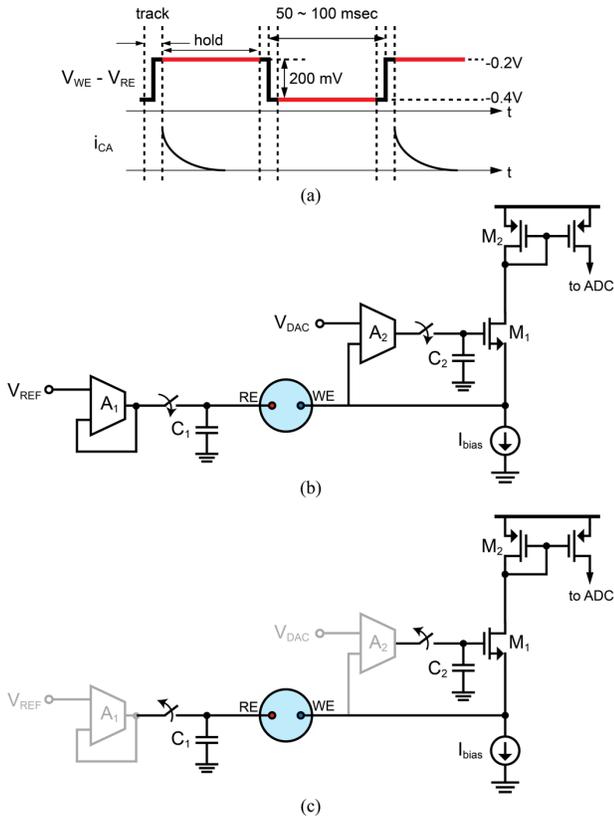

Fig. 8. (a) Proposed CA with sample-and-hold operation. (b) Simplified circuit schematics in the potential-tracking phase. (c) Simplified circuit schematics during the hold phase.

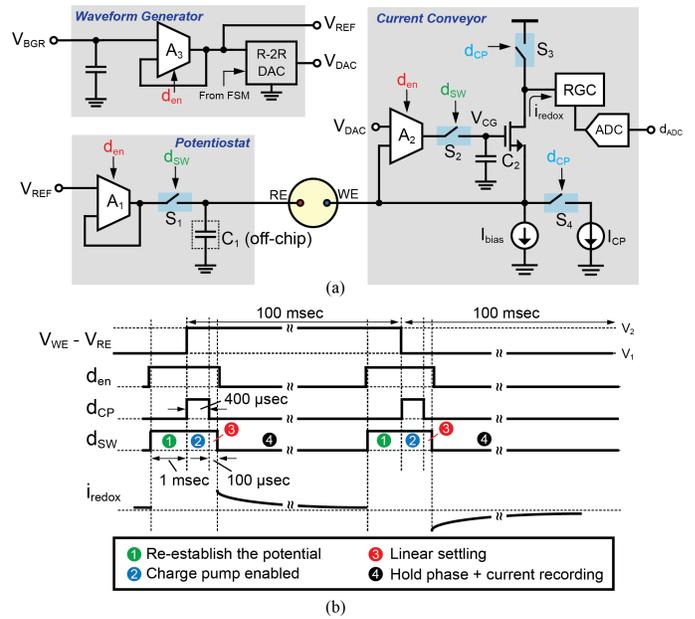

Fig. 9. (a) Complete circuit schematics. (b) Operational timing diagram.

*C. Sample-and-Hold Sensor Architecture*

Fig. 8 demonstrates the conceptual operation of the proposed sample-and-hold (S/H) electrochemical interfacing circuit. The central idea is to sample and hold the settled potentials onto capacitors $C_1$ and $C_2$ presented at both the reference (RE) and working electrodes (WE) using two switches ($S_1$ and $S_2$) such that the noise-contributing amplifiers can be disconnected and powered down during redox-current recording. Fig. 8(b) and 8(c) show the two modes of the operation. During the recording phase, the input-referred noise (IRN) current is only dominated by the common-gate device ($M_1$) and its mirroring load ($M_2$) and is insensitive to $C_{DL}$, leading to superior low noise at low-power operation. In this way, we have essentially eliminated the noise from Eq. (4). The degree of power reduction from this duty-cycling method is determined by the duration ratio between the tracking and the holding phases. Inevitably, there will be sampling noise including those from the amplifiers, charge injection, clock feedthrough, and environmental electromagnetic interference (EMI). This will cause undesirable modulation of the electrode potentials which affect the kinetics of the electron transfer, and thus the measurement uncertainty. As will be demonstrated later in Section IV-B, this can be compensated using principal component analysis (PCA) based on the multiple parameters extracted from the measurements.

In the design, $C_1$ (1 μF, an off-chip ceramic capacitor) is chosen to supply a sufficient amount of charge for the redox reaction without much perturbation to the electrode potential (~1mV under complete redox reaction) while $C_2$ (40 pF, on-chip MOSCap) is selected to minimize the sampling noise, charge injection, and clock feedthrough. As the detection is based upon the extraction of the time constant, the impact from both the offsets and the flicker noise is minimal. Therefore, mitigation techniques such as chopper stabilization and correlated double sampling are not incorporated. Due to its open-loop nature, the input impedance of the current buffer $M_1$ is elevated during the recording phase. In turn, this will cause drifts in the WE potential by the modulation from the time-varying redox currents. Therefore, it is critical to supply a decent amount of bias current (2 μA) to keep such a modulation within an acceptable margin. The simulated input impedance is ~16 kΩ at the typical corner, and the WE potential deviates no more than ~3.3-mV with ±200nA of current change.

*D. Circuits Implementation*

Fig. 9 shows the block diagram, the timing diagram, and the schematics of the proposed S/H electrochemical interfacing IC. The operation can be separated into four phases. Prior to each potential stepping, all the amplifiers are enabled and $S_1$ and $S_2$ are closed for 1 msec to re-establish the potentials. Upon ±0.2-V potential stepping, two charge pump switches ($S_3$ and $S_4$) embedded inside the WE current conveyor are closed for 400 μsec for fast charging and discharging. Afterward, a 100-μsec linear settling period is allocated, followed by the opening of $S_1$ and $S_2$ to hold the potentials and to disable the amplifiers. When operated continuously, the amplifiers are duty-cycled at 1.5% (=1.5msec/100msec). We extract the kinetics of the electron transfer rate from the recorded current transients. Most of the signal-of-interest lies within the first 10 msec after the potential stepping, indicating that the ADC can be further powered off during the other 90 msec. Identical Miller-compensated two-stage amplifiers (Fig. 10(a)) with a tail bias of 320 μA are used in the implementation of $A_{1-3}$ except that the compensation RC network is removed for the RE driver ($A_2$). This guarantees the settling speed of the amplifiers under a 1-μF load, and also serves as benchmarking comparison when configured into a conventional electrochemical sensor with feedback enabled. Note that the power consumption of $A_1$ and $A_3$ can be further reduced in the proposed S/H scheme for further power saving. The redox current is amplified by the regulated cascode (RGC) current mirror with programmable mirror ratio (Fig. 10(b)) and is digitized by a dual-slope ADC (Fig. 10(c)). The full scale (0.5 ~ 2 µA) and the integrating capacitance (1b control) in the ADC, as well as the RGC mirror ratio (1:8) are made adjustable to expand the total dynamic range by nearly 40 dB. No explicit filtering stage is incorporated; instead, we rely on the integration from the ADC with a window size of 50 µsec and is thus sub-optimal. To minimize leakage current during the hold phase, $S_1$ and $S_2$ are implemented with tri-switches and an analog keeper using a feedback unity-gain buffer (rail-to-rail input folded cascode amplifier, Fig. 10(d)). In this prototype study, the sensing IC is clocked at 30 MHz with a signal generator for testing flexibility. The programmability is offered by on-chip scanning registers and finite-state machine (FSM), and is controlled with an off-chip FPGA.

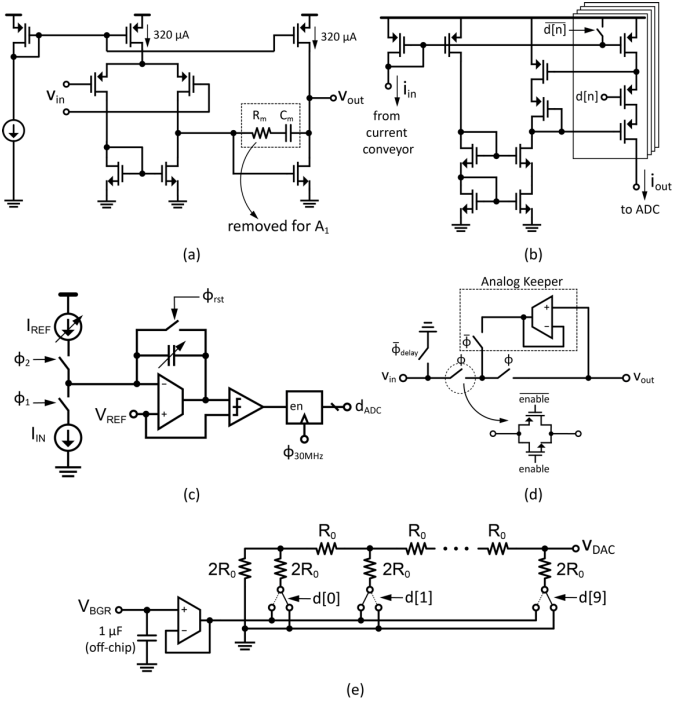

Fig. 10. Circuit schematics: (a) OTAs. (b) RGC. (c) Dual-slope ADC. (d) Tri-switch. (e) R-2R DAC.

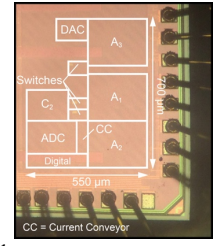

Fig. 11. Circuit micrograph.

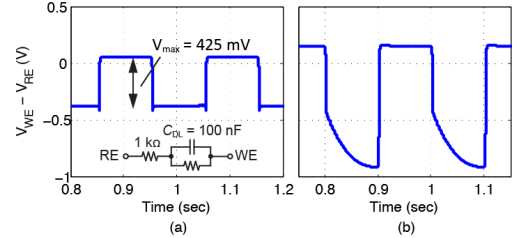

Fig. 12. (a) Measured potentials

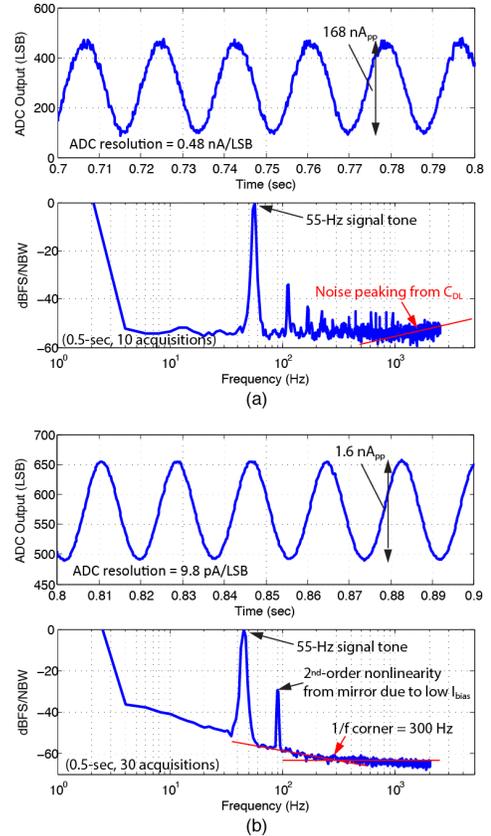

Fig. 13. Measured time-domain waveforms and PSDs with S/H: (a) disabled and (b) enabled.

## IV. Experimental Results

### A. Electrical Characterization

The prototype circuit is implemented in TSMC 65-nm CMOS technology. Fig. 11 shows the circuit micrograph with the core occupying an area of 0.385 mm$^2$, mainly dominating by the area of the three OTAs ($A_{1-3}$). All analog blocks are implemented with thick-oxide transistors and is powered from a 2.5-V supply through on-chip LDO. The digital circuits are operated at 0.9 V. To mitigate EMI, the circuit is powered using battery at 3 V, and the whole system is shielded with a Faraday cage except the FPGA and the clock source.

We first perform electrical characterization of the interface circuits. Specifically, we evaluate the settling behavior of the WE and RE potentials under abrupt changes. With the electrode impedance model (at a $C_{DL}$ of 100 nF) shown in Fig. 12(a), $V_{WE} - V_{RE}$ achieves complete settling at a maximum stepping amplitude of 425mV. This is 2× the desired amplitude (200 mV) given an overestimated $C_{DL}$, and thus it is sufficient to cover the variation in the sensor surface area. Fig. 12(b) also exhibits an example case when incomplete settling is also observed.

Next, we compare the sensor noise performance with and without the S/H technique. To do so, a sinusoidal voltage is injected through a 1-MΩ off-chip resistor and the samples are acquired during the hold phases at an extended time (0.5 sec). Fig. 13 compares the measurement results. Without the S/H, high-frequency noise-peaking from the 100-nF $C_{DL}$ is observable and the integrated noise over 2.5 kHz is 4.36 nA$_{rms}$.

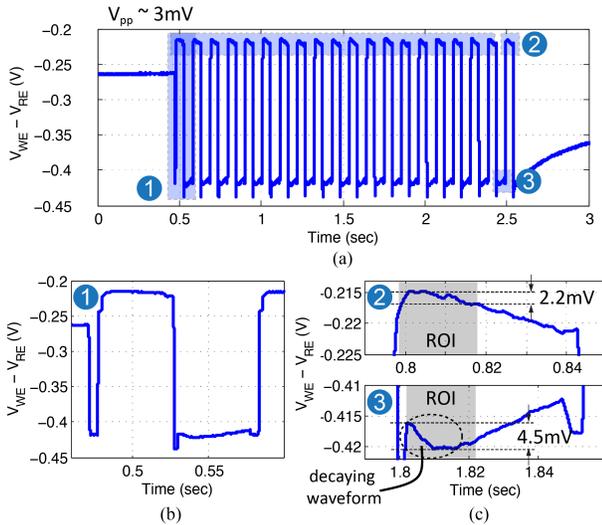

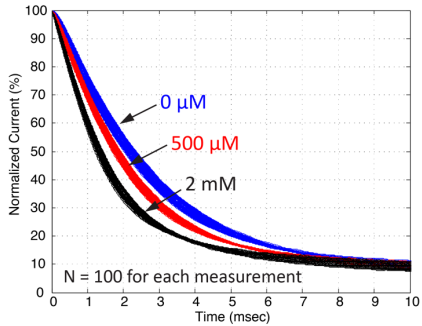

Fig. 14. (a) Measured electrode waveforms when interfacing with an aptamer device. (b) A zoom-in of the first pulse. (c) Waveforms during the hold phases.

Fig. 15. Measured normalized current responses.

On the other hand, a minimum IRN of 15.2 pA$_{rms}$ is measured with S/H enabled and stays insensitive to different $C_{DL}$ loading. The measured 1/f corner-frequency is ~300Hz. In these electrical testing, the mirror gain and the ADC parameters are adjusted accordingly to ensure low quantization noise. When interfacing with the aptamer sensors, these parameters are re-adjusted according to the measured current level. The maximum current that the current conveyor can measure is 1.6 µA$_{pp}$ with HD2 of 45 dB. The instantaneous dynamic range (DR) is about 60dB, which is sufficient in our application. The cross-range DR is ~100 dB (= 1.6 µA$_{pp}$ / 15.2 pA$_{rms}$).

The 10-bit R-2R DAC has a DNL and INL of +0.9/-0.58 and +1.2/-0.65 LSB, respectively. Based on our assay experiments, the requirement in the absolute accuracy is not as stringent so long as it is consistent across measurements. The circuit consumes 5.25 mW at 2.5 V without the S/H technique. Once S/H is enabled, all the power-hungry OTAs (but not ADC) are duty-cycled, reducing the total power to 0.22 mW.

### B. Assay Experiments

We prepare a two-electrode sensing probe functionalized with aminoglycoside aptamers and carry out the assay in target-spiked human whole-blood samples *in vitro*. The immobilization steps are similar to those in [15] and are summarized as follows: (1) define the sensor surface area on the gold electrode using heat-shrink tubing; (2) perform electrochemical cleaning in 0.5- and 0.05-M H$_2$SO$_4$ solution using three CV scans at 100 mV/sec from -0.4 to 1.5 V; (3)

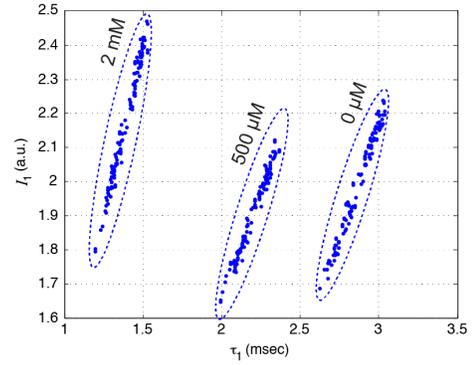

Fig. 16. Extracted $\tau_1$ and $I_1$.

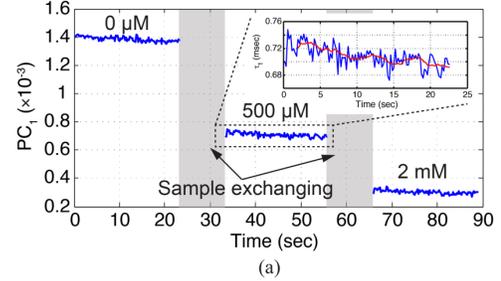

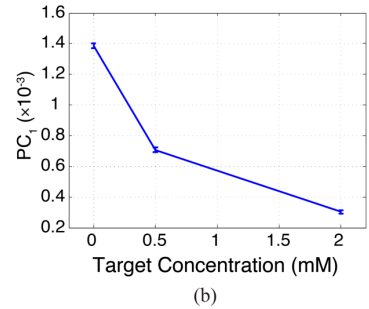

Fig. 17. Extracted principal component (a) vs. time and (b) vs. concentration. Error bar stands for one standard deviation.

Table III. Performance Comparison

| Electronics | Voltammetry Mode | Acquisition Time per Sample | Energy Consumption per Sample | Detection Limit (SNR = 1) |
|---|---|---|---|---|
| Palmsens EmStat3 | SWV with KDM | 8 sec (2 SWV scans) | n.a. | 1.5 µM |
| CMOS [25] | SWV with KDM | 8 sec (2 SWV scans) | 50 mJ | 18 µM |
| CMOS [This work] | CA + S/H | 0.1 sec | 22 µJ | 12.3 µM |
| CMOS [This work] | CA + S/H with average (N = 10) | 1 sec | 0.22 mJ | 3.1 µM |

incubate the sensor surface with 1-µM aptamer solution in saline sodium citrate (1× SSC) buffer for an hour at room temperature; (4) passivate the sensor surface with 6-mercapto-1-hexanol (MCH) at 6 mM in DI-water for two hours at room temperature; (5) store the device in 1× SSC at 4°C overnight to stabilize the SAM layer. Afterward, the device is ready to use and has a shelf life of one week. The human whole blood samples are purchased from BioIVT Inc. and are spiked with 200-mM kanamycin monosulfate (prepared in 1× SSC buffer) followed by serial dilution to the desired concentrations. To avoid sample contamination, we manually swap sample tubes

Table IV. Comparison Table

| | This work | | [25] | [28] | [27] | [33] | [35] | [36] |
|---|---|---|---|---|---|---|---|---|
| Applications | Personalized Pharmacokinetics | | | In vitro Diagnostics | | Neuroscience | | |
| Targets | Small-molecule drugs | | | DNA | | Dopamine | | |
| Recognition element | Structure-switching aptamers | | | DNA | | None (limited to electroactive molecules) | | |
| Tech. | 65 nm | | 65 nm | 250 nm | 130 nm | 65 nm | 250 nm | 65 nm |
| Voltammetry | SWV | Proposed S/H-CA | SWV | CV | CV | FSCV | FSCV/CV/CA | FSCV |
| Electrode, Area | Au, 0.25 mm$^2$ | | Au, 0.25 mm$^2$ | a-C, 1381 μm$^2$ | Au, 3025 μm$^2$ | Graphene, 1000 μm$^2$ | CFE/CNT-Au, 100 μm$^2$ | CFM, 2400 μm$^2$ |
| $C_{DL}$* | > 100 nF | | 100 nF | 167 pF | 800 pF | 1.1 nF | 1 nF | 1.5 nF |
| Potentiostat** | On-chip | | On-chip | Off-chip | On-chip | On-chip | n.a.$^{ss}$ | n.a.$^{ss}$ |
| Waveform Generator** | On-chip | | On-chip | Off-chip | On-Chip | Off-chip | Off-chip | Off-chip |
| Sensor IRN (SNR = 1) | 4.36 nA$_{rms}$ | 15.2 pA$_{rms}$[†] | 1.6 nA$_{rms}$ | 0.28 pA$_{rms}$[†] | 8.6 pA[†††] | 20 pA$_{rms}$ | 93; 21.6; 0.48 pA$_{rms}$ | 92 pA$_{rms}$ |
| Bandwidth | 2.5 kHz | 2.5 kHz | 2 kHz | 20 Hz | 1 kHz | 1 kHz | 10; 0.25; 0.11 kHz | 2 kHz |
| $I_{max}$ | ±800 nA | ±2.5 nA | ±800 nA | 12.5 nA | 350 nA | ±2.56 μA | ±50; ±50; ±0.2 nA | ±430 nA |
| Total DR ($I_{max}/I_{noise,rms}$) | 100 dB | | 60 dB[††] | 93 dB[††] | 93 dB | 108 dB | 104 dB | 79.4 dB |
| Electrochemical Data Acquisition Rate*** | 0.5 Hz[1] | 5 Hz[2] | 0.5 Hz[1] | 0.167 Hz[3] | 0.36 Hz[4] | 100 Hz[5] | 100 Hz[5] | 100 Hz[5] |
| Power | 5.25 mW | 0.22 mW | 6.64 mW | 0.25 mW$^c$ | 1.142 mW | 36 μW$^c$ | 12.1 μW$^c$ | 14.4 μW$^c$ |

*The noise performance is dependent on the electrical double layer capacitance in the electrochemical cell.
**Potentiostat is also termed "control amplifier" or "counter amplifier". Both the potentiostat and waveform generator (voltage DAC) inject extra noise and consume additional power and should be highlighted when comparing performance.
***This is referred to the time needed to acquire one electrochemical signal and is dependent on the voltammetry parameters.
[†]Measured lowest possible IRN.   [††]Instantaneous dynamic range.   [†††]ADC resolution limited.
$^c$Current readout circuit only (per channel).   $^{ss}$Two-electrode measurements with reference electrode connected to ground.
[1]SWV: -0.5V ~ -0.1V, 1mV per step, 60/400 Hz, with kinetics differential measurements.   [2]CA: -0.4 ~ -0.2V, 0.2V per step.   [3]CV: ±0.75V, 0.5 V/sec.
[4]CV: -0.7 ~ 0V, 0.5 V/sec.   [5]FSCV, -0.4 ~ 1.3V, 400 V/sec.

and perform chronoamperometry testing. We record $V_{WE}$ and $V_{RE}$ as well as the current transients for analysis.

Fig. 14 presents the measured $V_{WE} - V_{RE}$ waveforms with the actual aptamer-immobilized sensing device to study its settling and drooping. This is particularly important as the cell impedance critically depends on the integrity and the quality of the self-assembly monolayer (SAM). From Fig. 14(c), voltage changes in the region-of-interest (ROI, the first 10 msec after potential stepping) is ~ 4.5 mV, and its shape is correlated to the exponentially changing current transients. The drooping over a half-cycle duration (50 msec) is ~0.2 V/sec, indicating there exists a constant background leakage current of ~2 nA (assuming $C_{DL}$ of 10 nF) and can be viewed as the "offset" current caused by the pinholes in the SAM. A 3-mV$_{pp}$ sampling error in $V_{WE} - V_{RE}$ is observed from 20 consecutive recorded pulses, as shown in Fig. 14(a). This is much higher than the simulated sampling noise from the circuits alone and is therefore attributed to the finite pick-up from EMI effects. Such a sampling noise indeed affects the extracted kinetics and we will present a compensation technique in the next paragraph.

Fig. 15 shows the measured kinetics from *normalized* current transients at 0, 0.5, and 2 mM kanamycin concentration. Difference in the electron transfer kinetics at different target concentrations is observed. Next, we fit the *un-normalized* current transients using a two-term exponential function shown below to extract both the kinetics ($\tau_1$ and $\tau_2$) and the absolute current level ($I_1$ and $I_2$):

$$i(t) = I_1 e^{-\frac{t}{\tau_1}} + I_2 e^{-\frac{t}{\tau_2}}. \qquad (9)$$

Fig. 16 plots the extracted $\tau_1$ and $I_1$ at different concentration levels. Two observations are made from this figure. First, at each concentration, there exists high uncertainty in $\tau_1$. The extracted limit-of-detection (LoD, defined as SNR = 1) is 57 μM (11% of $K_D$). We attribute this to the random modulation in the electrode potentials during the S/H operation. In other words, in each chronoamperometric cycle, the perturbation on the held potentials causes slight modulation in the electron transfer kinetics, as described by the Butler-Volmer model [41]. Second, there exists a strong correlation between $\tau_1$ and $I_1$, and we believe $I_1$ serves as an indirect indicator for the sampled electrode potential. By taking advantage of this, we perform principal component analysis (PCA) to effectively compensate for the measured uncertainty. Fig. 17(a) plots the time series of the principal components after rotating the data. A significant reduction in the noise variance is demonstrated. Fig. 17(b) plots the principal components (mean and standard deviation) at different concentrations. An improvement in limit-of-detection from 57 to 12.3 μM is achieved. As chronoamperometry offers high temporal resolution (5 ~ 10 Hz), we further perform 10-point boxcar averaging to understand its detection limit and to compare with those measured using SWV. An LoD of 3.1 μM is measured at an acquisition rate of 1 Hz with drifts removed. Table III compares the molecular resolution acquired from different electronics and voltammetry techniques and Table IV provides a comparison with the state-of-the-art electrochemical sensing circuits. The use of S/H technique reduces both the circuit noise and the power consumption by a factor of 286× and 23×, respectively, increases the temporal resolution by 80× (or 8× with averaging), and improves LoD by 1.45× (or 5.8×).

V. CONCLUSION

In this work, we present the design and the analysis of an electrochemical circuit whose input-referred current noise is insensitive to the electrode capacitive loading to enable low-power monitoring of therapeutic drugs using redox-labeled structure-switching aptamers. We take advantage of the waveform simplicity in chronoamperometry and incorporate sample-and-hold operation around the electrochemical cell to store the desired electrode potentials using noiseless capacitors during the current-recording phase. In the proposed solution, the noise dominating OTAs are disconnected from the sensor and can be disabled, and simultaneous low-power and low-noise recording is achieved. We recorded an IRN of 15.2 pA$_{rms}$, and the performance is insensitive to the size of the cell-capacitance. We also addressed the issue of sampling error on the held potential using PCA-based signal processing. Using these strategies, we achieved an LoD of 3.1 μM in kanamycin concentration using an aminoglycoside aptamer with a $K_D$ of 0.5 mM. This LoD is > 5× better than those measured using standard SWV (without S/H) yet showing ~23× lower power consumption. Given that aptamer switches can be synthesized for a wide range of molecules, we believe our aptamer/CMOS platform could be used for many applications in healthcare. One exciting direction is the use of our S/H CA sensor for wearable or implantable systems. This will require strategies for mitigating the effect of EMI including shielding of the sensing probe or differential measurements using multiple channels. These topics are potential avenues for future research.

ACKNOWLEDGMENT

The authors would like to thank TSMC University Shuttle Program, Mentor Graphics, Dr. Peter L. Mage, Vlad Kesler, Nico Maganzini for the insightful discussions, and Ahmed Sawaby for the help in the CAD tool.